\iftrue 
\documentstyle[aps,prl,tighten,multicol,epsf
 ]{revtex} 
\def\widetext{\end{multicols} \noindent\hrulefill\vrule height.4em\hskip\columnsep\hfill\ }  
\advance\textheight by 0.2in
\def\narrowtext{ \hfill\hskip\columnsep\vrule depth.4em\hrulefill\begin{multicols}{2}\noindent}
\else\documentstyle[aps,prl,tighten,epsf]{revtex}
\fi

\def\fd{{\triangle}}            
\def\bd{\widetilde\triangle}    
\def\dmu{\partial_{\mu}}        
\def\dnu{\partial_{\nu}}        
\def\di{\partial_i}             
             
\def\dk{{d^3k\over(2\pi)^3}}    
\def\drho{\partial_{\rho}}      
\def\amu{a_{\mu}}               
                  
\def\arho{a_{\rho}}             
\def\bmu{b_{\mu}}               
\def\bnu{b_{\nu}}               
               
\def\brho{b_{\rho}}

\def\F{{\cal L}}                
\def\e{{\varepsilon}^{\mu\nu\rho}}          
\def\eij{{\varepsilon}^{ij}}

\def\K{{\cal K}}

\def\I{\hat{\rm I}}

\begin{document}

\title{{\hfill\large SU-ITP \#95/27, cond-mat/9511140\bigskip\\}
  Duality and Universality for the Chern-Simons bosons.}
\author{Leonid P. Pryadko\thanks{e-mail address: \tt
    leonid@quantum.stanford.edu} and 
  Shou-Cheng Zhang}  
\address{Department of Physics, Stanford University, Stanford, CA
  94305} 
\date\today

\maketitle
\begin{abstract}
  By mapping the relativistic version of the
  Chern-Simons-Landau-Ginzburg theory in 2+1 dimensions to the 3D
  lattice Villain x-y model coupled with the Chern-Simons gauge field,
  we investigate phase transitions of Chern-Simons bosons in the limit
  of strong coupling.  We construct algebraically exact duality and
  flux attachment transformations of the lattice theories,
  corresponding to analogous transformations in the continuum limit.
  These transformations are used to convert the model with arbitrary
  fractional Chern-Simons coefficient $\alpha$ to a model with
  $\alpha$ either zero or one.  Depending on this final value of
  $\alpha$, the phase transition in the original model is either in
  the universality class of 3D x-y model or a ``fermionic''
  universality class, unless the irrelevant corrections of cubic and
  higher power in momenta render the transition of the first order.
\end{abstract}
\begin{multicols}{2}

\section*{Introduction}

As one changes the external magnetic field, the Hall conductance of a
system of two-dimensional electrons jumps from one quantized value to
another, while the longitudinal conductance displays a peak. The zero
temperature localization length diverges in transition points,
indicating the second 
order quantum phase transitions.  Remarkably,
experiments\cite{Wei-88,Engel-90,Koch-91A} show that these phase
transitions are universal, {\em i.e.}\ the critical exponents
governing the divergence of the localization length are the same for
integer and fractional transitions.  It has also been
found\cite{Shahar-95A} that the critical resistances for integer and
fractional phase transitions are universal as well.  Finally, a
brilliant recent experiment by Shahar {\em et al\/}\cite{Shahar-95B}
demonstrates precise duality mapping of non-linear current-voltage
characteristics between the insulating and the fractional quantum Hall
sides of the transition, where carriers are respectively electrons and
fractionally charged quasiparticles.  One of the most challenging
problems in the quantum Hall effect is to understand the universality
of the transitions.

Apparently, there is a conceptual difference between the physics of
the integer and the fractional quantum Hall regimes.  In the former
case the quasiparticles are fermions, their localization transition is
believed to be a one-particle problem.  The physics is understood
quasiclassically as quantum tunneling-assisted
percolation\cite{Milnikov-88,Chalker-94,Dung-Hai-94}, while the
non-linear sigma model with topological term~\cite{Pruisken-88}
provides the field-theoretical formalism to study this problem.  On
the other hand, the excitations above the FQHE ground state, given
with extreme precision by inherently many-body Laughlin's
wavefunction\cite{Laughlin-83A}, have the infinite-range statistical
interaction that seem to invalidate the usual one-particle localization
theory.  Field-theoretically, this interaction can be described in
terms of the Chern-Simons gauge field coupled to either
bosons\cite{Zhang-89,Zhang-92}, or
fermions\cite{Jain-89,Lopez-91,Kalmeyer-92,Halperin-93A}.

Using the bosonic description, called the Chern-Simons-Landau-Ginzburg
(CSLG) theory\cite{Zhang-92} of the quantum Hall effect, Kivelson, Lee
and Zhang (KLZ) proposed\cite{Kivelson-92} a global phase diagram of
the quantum Hall effect.  The original problem of interacting
electrons in large magnetic field is mapped to the problem of bosons
in zero or weak uniform magnetic field, interacting via the
Chern-Simons gauge field.  The information about the filling factor is
contained only in the coefficient of the Chern-Simons term.  The
electromagnetic response functions in the CSLG theory can be expressed
solely in terms of response functions of the bosonic field and,
therefore, the different quantum Hall transitions are in the same
universality class if the critical exponents of the boson superfluid
to insulator phase transition are independent of the Chern-Simons coefficient.

On the basis of the CSLG theory in the RPA approximation KLZ argued
that the critical exponents of Hall transitions are universal; they
predicted values for critical conductances that have been recently
confirmed experimentally\cite{Shahar-95A}.  However, since the
Chern-Simons term is a marginal operator by na{\"\i}ve power counting,
it could in principle change the critical exponents.  If this happens,
the transition will no longer be universal.  Therefore, it is highly
desirable to check the idea of universality in some specific
field-theoretical calculation.  Unfortunately, it is very hard to
treat the full interacting problem in the presence of disorder.  For
this reason the analysis is often restricted to toy models that
nevertheless share some essential features with real 
quantum Hall systems.

The most spectacular feature of the quantum Hall system is the
symmetry of the phase diagram, and there are several field-theoretical
models with similar symmetry.  A broad class of such models has been
introduced by Shapere and Wilczek\cite{Shapere-89} who generalized the
construction of Cardy and Rabinovici\cite{Cardy-82A,Cardy-82B} and
constructed $4$ and $2$ dimensional self-dual models consisting of
mutually dual discrete Abelian gauge lattice models coupled via the
theta term.  The symmetry of similarly constructed $2+1$ dimensional
model with Chern-Simons term was analyzed by Rey and Zee\cite{Rey-91}.
L\"utken and Ross\cite{Lutken-92,Lutken-93} discussed the symmetry of
fixed points of self-dual $2$-dimensional clock model and argued that
the corresponding group $SL(2,Z)$ describes the symmetry of {all\/}
fixed points of the quantum Hall system.  This statement is analogous to
the Law of Corresponding states\cite{Kivelson-92}; unfortunately, so
far no derivation of the field-theoretical realization of this
symmetry group is known.

The question whether phase transitions in the system of Chern-Simons
bosons can even in principle be universal is so general that the
answer for 
virtually any such model would be interesting.
Particularly, one can study the phase transition driven by the
amplitude of external periodic potential at commensurate filling
instead of that driven by disorder.  It is known\cite{Fisher-89A} that
the system of bosons at fillings commensurate with external periodic
potential can be mapped to the x-y model.  In the quantum Hall problem
one arrives\cite{Wen-Wu-93} at the x-y model minimally coupled with
the Chern-Simons term
in zero average magnetic field.
Within this model one can naturally address the question whether the
order-disorder transition is altered by coupling with the Chern-Simons
term and, if so, whether there is a new universality class, or the
phase transition continuously depends on the Chern-Simons coupling.
Even though this is an artificial model, its successful analysis will
undoubtedly shed some light on the universality of the real quantum
Hall phase transitions.

Wen and Wu\cite{Wen-Wu-93} used the representation of the the x-y
model in terms of the relativistic complex-valued scalar field with
quartic interaction, especially convenient for perturbational
expansion.  They also used the $1/N$ expansion, known to work for
ordinary bosons, to regularize the perturbational series in three
dimensions.  Within the second loop approximation Wen and Wu found
that although the phase transition remains of the second order, the
critical exponents of the x-y model are continuously changed by the
Chern-Simons coupling.
 
Soon after that, we pointed out\cite{Pryadko-94} that the $1/N$
expansion artificially 
suppresses the gauge fluctuations in this model, and the result of Wen
and Wu is not likely to hold in the physical limit of $N=1$.  Our
alternative calculation\cite{Pryadko-94} was performed in the vicinity
of nominal tri-critical point of the theory, where the quartic
interaction is absent and one has to keep the six-field coupling.  The
effective theory in the vicinity of the tricritical point is
renormalizable in three dimensions and allows systematic loop
expansion, exactly as the theory with four-field interaction is
renormalizable in four dimensions where the quartic coupling is
dimensionless.   The presence of the Chern-Simons coupling, also
dimensionless in 3D, 
does not introduce any classically divergent quantities, but it has a
dramatic effect already at the second loop level, changing the
transition from the second to the first order.  This is similar to the
statement\cite{Halperin-74} that the phase
transition in clean type I superconductors becomes weakly 
first order due to the fluctuations of the electromagnetic field. 
In our case, since the theory without the quartic
interaction is massive, it remains massive in some finite region of
values of the quartic coupling.  Thus, within the perturbative region of
the relativistic complex scalar field model, the transition is of the
first order even for some non-zero values of the quartic coupling.

This conclusion obtained in the vicinity of the Gaussian fixed point
of the scalar sector was not in formal contradiction with the
result of Wen and Wu obtained near the strongly coupled fixed
point (formally accessed with the $1/N$ expansion,) but it reopened the
question about the universality of phase transitions of
scalar fields coupled with the
Chern-Simons field.  The results of both calculations indicate that
the Chern-Simons 
interaction is a {\em relevant\/} perturbation to the scalar theory
with strong short-range repulsion 
and some kind of non-perturbative analysis is necessary.

The goal of this paper is to address the question of universality of
phase transitions in relativistic Chern-Simons bosons without the
limitations of the perturbation theory.  We access the strong coupling
limit of the Chern-Simons bosons using the Villain form of the x-y
model minimally coupled with the Chern-Simons gauge field.  For this
model we formulate the exact duality and flux attachment {\em
  transformations\/} with usual properties in the continuum limit and
build the Haldane-Halperin\cite{Haldane-83,Halperin-84a} hierarchy of
models related by these non-local transformations.  We show that in
the absence of external magnetic field the universality class of phase
transition in the strong-coupling limit is determined by the
Chern-Simons coupling $\alpha=p/q$.  When either $p$ or $q$ is even,
the phase transition is in the universality class of the
order-disorder transition in the usual three dimensional x-y model.
When both $p$ and $q$ are odd, the system in the vicinity of critical
point (if any) is equivalent to x-y model with the Chern-Simons
coupling $\alpha=1$, corresponding to the Fermi statistics.  We
believe that our result is exact and apply at least for fractions with
small enough denominators $q$.  With increasing denominators the terms
formally 
irrelevant near the x-y critical point grow,
effectively limiting the number of fractions with expected
universality.

Unlike the previous works\cite{Frohlich-89,Luscher-89,%
  Muller-90,Fradkin-89A,Fradkin-90,Eliezer-92C} analyzing lattice
Chern-Simons models, we derive algebraically exact flux attachment
and duality transformations.  Thus we avoid the problem of relevance
or irrelevance of the higher order in momenta terms
appearing in different lattice definitions of the Chern-Simons
coupling.  Instead, we construct the exact transformation to the
theory with zero $\alpha$, equivalent to the usual 3D x-y model with
finite lattice temperature.   For this theory the algebra of local critical
operators is known, and the formal irrelevance of extra terms does not
require additional proof.

The paper is organized as follows.  We derive the exact duality
transformation for the Villain form of the x-y model in the
Section~\ref{sec:x-y-duality} 
and the exact flux attachment transformation in the
Section~\ref{sec:periodicity}. 
These transformations have correct na\"{\i}ve continuum limit and work for
lattice theories with gauge coupling of most general form.  In the third
section we construct the sequence of such transformations leading to
Villain x-y model with the Chern-Simons coupling either zero or one
and utilize the known properties of the usual x-y model to build the
phase diagram of the original model.

\section{Generalized duality}
\label{sec:x-y-duality}
Particles in superfluid repel each other at short distances.  At very
small temperatures most of the particles are in the same quantum
state, characterized by the condensate wavefunction
$\Psi(x,t)=\psi\exp i\theta$, where both the amplitude $\psi$ and the
phase $\theta$ are real-valued functions.  Slow phase rotations have
extremely small energy, implying the existence of a linear sound mode.

Another important kind of excitations in superfluid are vortices in two
dimensions, or vortex lines in three dimensions.  They are
characterized by non-trivial phase of the condensate wavefunction
gained along any surrounding contour.  Far enough from the center of
the vortex the gradient of phase is small and the superfluid
density $\psi^2$ is close to its non-perturbed value.  Although the
reduced 
density in the core region near the center of the vortex costs some
energy, the total energy of a single vortex is mostly determined
by the twisted phase in the area outside the core and any two
vortices have a long range logarithmic interaction. 

Quite differently, vortices in superconductors have only short-range
interactions (just like {\em particles\/} in superfluid): the extra
phase is easily screened by the vector potential.  Instead, the
conserved particle currents in normal phase interact 
logarithmically through their magnetic fields just like {\em vortices\/}
in superfluid.  This analogy can be formulated more precisely as the
{\em duality\/} between the type II superconductors in London limit
where the penetration length of the magnetic field is large compared
to the coherence length and the strong-coupling limit of the
Ginzburg-Landau model.  In this limit of strong coupling one can
completely neglect the 
density fluctuations, considering only the phases $\theta_n$ as 
the only degrees of freedom.  The core energy of the vortices can be
regularized by redefining the theory at the lattice.  The resulting
x-y model is a collection of classical spins
$s_n=(\cos\theta_n,\sin\theta_n)$ with nearest-neighbor interaction of
the form ${\bf s}_i {\bf s}_j=\cos(\theta_i-\theta_j)$.  It turns out
that the reverse transformation from the x-y model to the strongly
coupled superfluid is also possible\cite{Kleinert-82}.  This two-way
analogy has been confirmed by comparing the results of numeric
computations in the lattice x-y model with corresponding analytical
results for the $\phi^4$ theory.

The partition function of the x-y model 
\begin{equation}
  \label{xy-partition}
  Z_{\rm x-y}= \int{d\theta_n\over2\pi} \exp{\sum
    {\cos(\theta_{n+\mu}-\theta_n)\over T}},
\end{equation}
is periodic with respect to the phases $\theta_n$ and, at least in the
strong coupling limit where the lattice temperature\cite{Note-1}~$T$ 
is small, is mainly determined by the vicinity
of maxima of the expression in the exponent.  The shape of these
maxima can be simulated using the Villain form of the x-y model 
$$ %
\exp{{\cos(\theta_{n+\mu}-\theta_n)}\over{}T}
\longrightarrow\sum_m
  \exp{\displaystyle 
-{(\theta_{n+\hat\mu}\!-\!\theta_n\!-\!2\pi{}m_{n\hat\mu})^2\over2T}} 
$$ %
originally proposed by Berezinskii\cite{Berezinskii-71} and later
independently and in greater details investigated by
Villain\cite{Villain-75}.

Peskin\cite{Peskin-78} proved that the 
duality transforms the Villain 
x-y model into the so-called frozen superconductor, or the zero
temperature limit of the Villain x-y model coupled to the Maxwell
field.  This rigorous analysis has been extended by
Kleinert\cite{Kleinert-82} who analyzed the lattice superconductor in
wide region of parameters using the Villain approximation and
achieved quantitative agreement\cite{Kleinert-89} with numerical
simulations.  These model computations eventually
lead\cite{Fisher-89,Kovner-91,Kiometzis-93} to 
understanding of the complete phase diagram of superconductors in
the dual representation.  Later Lee and  
Fisher\cite{Lee-89,Dung-Hai-91} and Lee and Zhang\cite{Lee-Zhang-91}  
applied this duality transformation to the anyon
superconductivity and the quantum Hall effect. 

We are interested in a very similar class of models, namely the strong
coupling limit of relativistic Chern-Simons bosons.  The idea is to
follow the successful example of superconductor and examine
non-perturbative properties of this model at the lattice using the
Villain form of x-y model coupled to the Chern-Simons gauge field.   
There are several\cite{Frohlich-88,Luscher-89,Muller-90,Eliezer-92C}
different gauge-invariant definitions of the lattice Chern-Simons
coupling, and, in order to keep the analysis as universal as possible,
we consider the three-dimensional Villain x-y model
\begin{eqnarray}
  \label{villain-xy}
  &\relax & {Z[A]\!=\!\prod_{n\mu}\!
    \int\limits_{-\pi}^{\pi}\!{{d\theta_n\over2\pi}\!\!\!\!
      \sum_{m=-\infty}^\infty\!\!  \exp-{\left(\fd_{\mu}
          \theta_n\!-\!A_{n\mu}\!-\!2\pi
          m_{n\mu}\right)^2\over2T}},}\\
  \label{gauge-part}
  &\relax & Z\!=\!\int{\exp \left(-{1\over2} 
      A_{n\mu}\K_{nn'}^{\mu\nu}
      A_{n'\nu}\right)Z[A]\prod_{n,\rho}d A_{n\rho}},
\end{eqnarray}
at the cubic lattice, minimally coupled to the gauge field $A_{n\mu}$
with some generic, possibly non-local kernel $\K_{nn'}^{\mu\nu}$,
or $\K^{\mu\nu}({\bf k})=\K({\bf k})$ in momentum
representation.  By the {\em generic Chern-Simons
coupling\/} we shall imply that in the continuum limit $\K$ must be
an antisymmetric matrix linear in momenta.  

The summation over links $\mu$ in~(\ref{villain-xy},\ref{gauge-part})
is performed only in the positive direction $\hat\mu$; the phases
$\theta_n$ are defined in the nodes, while both the integer-valued
field $m_{n\mu}$ and the gauge field $A_{n\mu}$ are defined at the
links of three-dimensional cubic lattice.  We use the notations
$\fd_{\mu}f_n=f_{n+\hat\mu}-f_n$, $\bd_{\mu}f_n=f_{n}-f_{n-\hat\mu}$
for the forward and backward lattice differences respectively, and
imply the summation over repeated indices.  

It will be convenient to specify the Fourier expansion of vector
fields in the form
\begin{equation}
  \label{fourier-decomposition}
  A_{n\mu} = {1\over\sqrt{N}}
  \int\!\!\!\!\!\int\limits_{-\pi}^{\pi}\!\!\!\!\!\int{%
    d^3k\over(2\pi)^3}{%
    A_{\mu}({\bf k})e^{\textstyle ik_{\!\mu}\!/\!2} e^{\textstyle
      i{\bf k r}_n}}
\end{equation}
(no summation in $\mu$!) 
to emphasize their location at the links of the lattice.  With this
definition the gradient $\fd_{\mu}f_n=\bd_{\mu}f_{n+\hat\mu}$ or
$P_{\mu}f({\bf k})$ in momentum representation is a vector, while the
divergence $\bd_{\mu}A_{n\mu}=\fd_{\mu}A_{n-\hat\mu\mu}$ or
$P_{\mu}A_{\mu}({\bf k})$ is a scalar field, as one would expect from
the geometrical interpretation of the lattice fields.  We shall use the
vector $P_{\mu}=2\sin(k_{\mu}/2)$ to define different
functions of lattice momentum.  For example, the kernel of the local
in the gauge field Chern-Simons term used in
refs.\cite{Frohlich-88,Luscher-89,Muller-90} can be written as
$\K_{\mu\nu}=\e P_{\rho}Q_0/2\pi\alpha$ with
$Q_0=\cos\sum_{\mu}k_{\mu}/2$, and in the
Section~\ref{sec:periodicity} we shall define the Chern-Simons 
term with the kernel $\K_{\mu\nu}=\e P_{\rho}/2\pi\alpha Q_0$
providing for the local coupling between the gauge-invariant scalar
currents on the lattice.  

Let us treat the action $Z[A]$ as the Villain x-y model in the presence
of some external gauge field $A$, and follow the
derivation\cite{Peskin-78} of the duality transformation for this
model.  While the integration in phases $\theta_n$ is performed over
restricted intervals, the gauge transformation
\begin{equation}
  \label{int-gauge-transformation}
  \begin{array}{rcl}
    \theta_n&\rightarrow&\theta_n+2\pi N_n\\
    m_{n\mu}&\rightarrow&{}m_{n\mu}+ N_{n+\hat\mu}- N_n
\end{array}
\end{equation}
does not change the integrand of (\ref{villain-xy}).  One can extend
the integration region in $\theta_n$ to the infinite interval by
simultaneously constraining the field $m_{n\mu}$ to avoid over
counting.  This can be done, for example, by applying the usual gauge
condition
\begin{equation}
  \label{int-gauge-constraint}
  0=\sum_{\mu}\left(m_{n+\hat\mu\mu}-m_{n\mu}\right).
\end{equation}
Now introduce an auxiliary field $b_{n\mu}$ by writing
\begin{eqnarray}
  \label{splitted}
  Z[A]&=& \prod_{n}\int_{-\infty}^{\infty}{d\theta_n\over2\pi}
  \prod_{\mu}\left({T\over2\pi}\right)^{1/2} \int
  db_{n\mu}\sum_{m_{n\mu}}\\
  &\relax&\hskip-0.8cm\times\exp\left[-{T\over2}b_{n\mu}^2
    +ib_{n\mu}\left(\fd_{\mu}\theta_{n} -A_{n\mu}
      -2\pi{}m_{n\mu}\right)\right].\nonumber
\end{eqnarray}
The integration by $\theta_n$ can be done after regrouping the terms
in the exponent,
\begin{equation}
  \label{conservation-of-b}
  \int_{-\infty}^{\infty} {d\theta_n\over2\pi}
  e^{\textstyle-i\theta_n\bd_{\mu}b_{n\mu}}=
  \delta\left(\bd_{\mu}b_{n\mu}\right),
\end{equation}
the resulting constraint being merely the lattice version of the
equation ${\bf \nabla\cdot b}=0$ indicating that the field $\bf b$ is
a pure curl,
\begin{equation}
  \label{lattice-curl}
  b_{n\mu} =\e\bd_{\nu}a_{n-\hat\rho\rho}.
\end{equation}

So far the only difference with the duality
trans\-for\-ma\-tion\cite{Peskin-78} for the pure x-y model is the
presence of the product $-ib_{n\mu}A_{n\mu}$ in the exponent
of~(\ref{splitted}).  This extra term in the action  is
responsible for important additional symmetry.  Indeed, the original
gauge field in the full partition function~(\ref{gauge-part}) can be
integrated away,
\begin{eqnarray}
  Z&\!=\!&\prod_{n\mu}\left({T\over2\pi}\right)^{1/2}
  \int_{\theta_n,b_{n\mu}} \sum_{m_{n\mu}} e^{\textstyle
    ib_{n\mu}\left(\fd_{\mu}\theta_n
      -2\pi{}m_{n\mu}\right)}\nonumber\\ %
  \label{integrated-over}
  &\times&\exp\left[-{1\over2} \int {d^3k\over(2\pi)^3} b_{\mu} (-{\bf
      k}) \left(T\delta_{\mu\nu}\! +\!\K^{-1}_{\mu\nu}\right) b_{\mu}({\bf
      k}) \right],
\end{eqnarray}
and, since the field $b$ is transverse, it is clear that the
partition function $Z$ depends only on the transverse combination
\begin{equation}
  \label{combination}
  T\delta^{\rm t}_{\mu\nu} +\K^{-1}_{\mu\nu},
\end{equation}
indicating the exact equivalence of any two models with the
parameters satisfying the relationship
\begin{equation}
  \label{equivalence-general}
  T_1\delta^{\rm t}_{\mu\nu} +{\K_1}^{-1}_{\mu\nu}
  =T_2\delta^{\rm t}_{\mu\nu}+{\K_2}^{-1}_{\mu\nu},
\end{equation}
where $\delta^{\rm t}_{\mu\nu}=
\delta_{\mu\nu}-\hat{P}_{\mu}\hat{P}_{\nu}$ denotes the
transverse part of the Kronecker symbol and 
$\hat{P}_{\nu}\equiv P_{\nu}/|P|$ is unit vector.  This
reparametrization changes both the gauge and scalar couplings; it is
an exact symmetry of the partition function of the Villain x-y model
or any correlators that do not involve the gauge field directly.
These are the only physical correlators as long as we treat the
Chern-Simons field as an auxiliary field 
needed to define the fractional statistics\cite{Wilczek-83} for
the particles.  
Since the lattice temperature $T$ measures the strength of the local
repulsion, the relationship~(\ref{equivalence-general}) implies that a
part of this repulsion can be mediated by the gauge field if its
propagator has non-zero symmetric part. 

One may further verify that the system averages taken in the presence
of any number of vortex-antivortex pairs
$\exp{i(\theta_{n}-\theta_{n'})}$, %
as well as the monopoles dual to them, do not depend on particular
selection of parametrization.  Moreover, although at the tree level
the properties of quasiparticles appear to be changed, any correlators
involving the gauge-invariant current retain their values.  This can
be checked by performing the same transformation in the
presence of additional external gauge field $A_0$: the
reparametrization does not affect the minimal gauge coupling with this
external field. 

Having established the reparametrization
symmetry~(\ref{equivalence-general}) of the 
complete model~(\ref{villain-xy},\ref{gauge-part}), let us resume the
transformations of the partition function~(\ref{splitted}).  As usual,
the field $a_{n\rho}$ introduced in~(\ref{lattice-curl}) is defined up
to a gauge transformation
\begin{equation}
  a_{n\mu}\rightarrow a_{n\mu}+(\lambda_{n+\mu}-\lambda_{n})
\end{equation}
with arbitrary $\lambda_n$; for simplicity we imply the Coulomb gauge
$\bd_{\mu} a_{n\mu}=0$ for all our gauge fields.  The integration over
$\theta_n$ results in
\begin{equation}
  Z[A]=\int_{a}\sum_m e^{\textstyle-T b_{n\mu}^2/2-2\pi i
    b_{n\mu}m_{n\mu}-i A_{n\mu}b_{n\mu} },
\end{equation}
or, after regrouping the second term in the exponent,
\begin{equation}
  \label{vort-sum}
  Z[A]=\int_{a}\sum_m e^{\textstyle- T b_{n\mu}^2/2+ 2\pi i
    a_{n\mu}M_{n\mu}-i A_{n\mu}b_{n\mu} },
\end{equation}
where the integer-valued vorticity is defined as
\begin{equation}
  M_{n\mu} =\e
  \left(m_{n+\hat\mu+\hat\nu\rho}-m_{n+\hat\mu\rho}\right).
\end{equation}
Since the vorticity is locally conserved,
\begin{equation}
  \label{vorticity-conserved}
  \sum_{\mu} \left(M_{n\mu}-M_{n-\mu\mu}\right)=0,
\end{equation}
we may exchange the summation over $m_{n\mu}$ for a sum over
$M_{n\mu}$ subject to condition~(\ref{vorticity-conserved}).  This
integer-valued constraint can in turn be removed by new
phases~$\theta_n$,
\begin{equation}
  \delta_{0,\bd_{\mu}M_{n\mu}}= \int\limits_{-\pi}^{\pi}
  {d\theta_n\over2\pi} e^{\textstyle i\theta_n\bd_{\mu}M_{n\mu}}.
\end{equation}
To finish the transformation of the dual model towards the Villain
form analogous to the original
model~(\ref{villain-xy},\ref{gauge-part}), Peskin\cite{Peskin-78}
introduced a convergence factor 
\begin{equation}\label{convergence-factor}
  1=\lim_{t\rightarrow0}\exp\left[-{t\over2}M_{n\mu}^2\right]
\end{equation}
into Eq.~(\ref{vort-sum}) and transformed the resulting sum
\begin{equation}
  \label{zero-dual-currents}
  \lim_{t\rightarrow0}\sum_{M}e^{\textstyle %
    -{t}M_{n\mu}^2/2 +2\pi i a_{n\mu}M_{n\mu}
    +i\theta_n\bd_{\mu}M_{n\mu}}
\end{equation}
with the Poisson summation formula
\begin{equation}
  \label{summation-formula}
  \sum_{M=-\infty}^{\infty}\!\!\!  e^{\textstyle i\varphi
    M\!-\!{t}M^2/2} \!=\!\sqrt{2\pi\over t}\!\!\!
  \sum_{m=-\infty}^{\infty}\!\!\!  e^{\textstyle-(\varphi\!-\!2\pi
    m)^2/2t}.
\end{equation}
After rescaling the gauge fields $2\pi a_{n\mu}=\tilde A_{n\mu}$, $2\pi
b_{n\mu}=\tilde B_{n\mu}$, the dual action can be finally written as
\widetext
\begin{equation}
  \label{villain-xy-dual}
  Z[A]=\lim_{t\rightarrow0}\prod_{n}
  \int\limits_{-\pi}^{\pi}{d\theta_n\over2\pi}
  \prod_{\mu}\int{dA_{n\mu}} \sum_{m} \exp\left[
    {-{1\over2t}\left(\fd_{\mu}\theta_{n}\!  -\!2\pi
        m_{n\mu}\!-\!\tilde{A}_{n\mu}\right)^2
      -{i\over2\pi}A_{n\mu}\tilde{B}_{n\mu}
      -{T\over8\pi^2}\tilde{B}_{\mu}^2}\right].
\end{equation}
Again, we can integrate away the original gauge field $A_{n\mu}$ to
obtain the final form of the dual action
\begin{equation}
  \label{dual-zero-temp}
  Z=\lim_{t\rightarrow0}\int_{\tilde{A},\theta}\sum_m \exp\left[
    -{1\over2t}\left(\fd_{\mu}\theta_{n}\! -\!2\pi
      m_{n\mu}\!-\!\tilde{A}_{n\mu}\right)^2
    +{1 \over8\pi^2}\tilde{A}_{-k} %
    {\bf P}\!\times\!\left(T\!+\!\K^{-1}\right)\!\times\!{\bf P} %
    \tilde{A}_{k} %
  \right],
\end{equation}
\narrowtext %
\noindent
where the appropriate summation in each term of the exponent is implied.
In accord with our previous conclusion, the action depends only on the
transverse part of the combination
$(\K^{-1}+T\delta^t)_{\mu\nu}$.  In addition, the dual model has a
form of (\ref{villain-xy},\ref{gauge-part}) with zero lattice
temperature $t=0$ and the dual gauge kernel\cite{Note-2}
\begin{equation}
  \label{zero-dual-kernel} 
  \tilde\K_0 %
  ={TP^2-{\bf P}\!\times\!\K^{-1}\!\times\!{\bf P}\over(2\pi)^2}
  ={P^2\over(2\pi)^2}\left(T+\K^{-1}\right),
\end{equation}
where only the transverse part enters because of the gauge
fixing for the field $\tilde A$.  The dual model has the same form as
the original one and, since the new gauge field was introduced as an
auxiliary field, this model has the same reparametrization
freedom~(\ref{equivalence-general}).  
Therefore, the most general form of the dual model with the lattice
temperature $\tilde T$ and the gauge kernel $\tilde\K$ satisfies the
equation 
\begin{equation}
  \label{zero-equiv}
  \tilde T+\tilde\K^{-1}=0+\tilde\K_0^{-1}={(2\pi)^2\over
    P^2}\left(\K^{-1}+T\right)^{-1},
\end{equation}
or, more symmetrically,
\begin{equation}
  \label{duality-general}
  (T+\K^{-1})(\tilde{T}+\tilde\K^{-1})={(2\pi)^2\over P^2}.
\end{equation}
An alternative direct derivation of the finite temperature dual
action, as well as the vortex---monopole mapping explicitly relating
the scalar sectors of the two models is provided in the
Appendix~\ref{duality-derivation}.  This algebraically exact mapping
proves that dual 
models of the form (\ref{villain-xy},\ref{gauge-part}) related by the
generalized duality~(\ref{duality-general}) are
equivalent, being just two different representations of the same
model.

To provide an example of 
the duality in the presence of the Chern-Simons coupling, let us
consider the simplest linear in momenta form of the gauge kernel 
\begin{equation}
  \label{non-local-cs}
  \K^{\mu\nu}({k})={1\over2\pi\alpha}\e P_{\rho}.
\end{equation}
While this definition is very well convergent towards the continuum
limit and generally looks like a plausible definition of the
Chern-Simons kernel, it is actually non-local in coordinate
representation.   The conventional duality transformation results in
zero lattice temperature model~(\ref{dual-zero-temp}) with the gauge
kernel 
\begin{equation}
  \label{frozen-dual-kernel}
  \tilde\K_0^{\mu\nu}(k)=-{\alpha\over2\pi}\e P_{\rho}
  +{T\over(2\pi)^2}\left[ P^2\delta^{\mu\nu}-P^{\mu}P^{\nu}\right],
\end{equation}
analogous to the frozen superconductor\cite{Peskin-78}.  The two
terms here are the non-local Chern-Simons term of the same form
as~(\ref{non-local-cs}) and the usual lattice Maxwell action.  The
zero lattice temperature implies zero tree level propagator for the
matter field, in analogy with the dual theory\cite{Zhang-92} of
non-relativistic continuum bosons coupled to the purely statistical
Chern-Simons gauge field.  However, already in the RPA approximation,
the dynamics of the dual gauge field $\tilde{A}$, associated with
non-zero dimensionful charge $e^2=(2\pi)^2/T$, renders this propagator
finite, therefore introducing some non-zero lattice temperature for
the dual matter field.   

Let us try to understand this 
result using our generalized duality
transformation~(\ref{duality-general}).  Specifically, we  
want to find a form of the dual model with purely statistical gauge
field, so that all dynamics would be associated with the scalar field.

To simplify the subsequent algebra, it is convenient to introduce
special notation for 
typical matrices appearing in the expressions.  The matrices
$\delta^{\rm t}_{\mu\nu}=\delta_{\mu\nu}- \hat{P}_{\mu}\hat{P}_{\nu}$
and $[\I]_{\mu\nu}=\e \hat{P}_{\rho}$ commute with each other, while
$[\I^2]_{\mu\nu}=-\delta^{\rm t}_{\mu\nu}$.   Algebraically, the
symmetric matrix 
$\delta^{\rm t}_{\mu\nu}$ is equivalent to unity, while $\I$ plays a
role of $i=\exp({i\pi/2})$. 
Since we always assume the transverse gauge, it is possible not
to write $\delta^{\rm t}_{\mu\nu}$ at all and treat $\I$ as a
commuting number.  In these notations the dual gauge
kernel~(\ref{frozen-dual-kernel}) can be written as
$$ %
\tilde K_0=-\I{\alpha P\over2\pi}+{T P^2\over(2\pi)^2},
$$ %
so that the corresponding inverse matrix in the transverse gauge is
just
\begin{eqnarray}
  \label{new-temp-deriv}
  {1\over \tilde K}+\tilde T \; &\stackrel{\textstyle\rm
    (\ref{zero-equiv})}{=}&\;{1\over\tilde K_0} ={(2\pi)^2\over-2\pi
    \I\alpha P +T P^2}
  \nonumber\\ %
  &\approx& {2\pi{\I}\over\alpha P }\left(1-{\I T P
      \over2\pi\alpha}\ldots\right)
        \nonumber\\ %
        & =& {2\pi \I\over\alpha P } +{T\over\alpha^2}+\ldots
\end{eqnarray}
Na\"\i vely, the first term of the expansion may be associated with
the inverse of the new Chern-Simons kernel, while the second constant
term plays a role of the dual temperature.  We notice that while the
{\em Chern-Simons term is the same as the one in the representation
  with $T=0$, the $F^2$ term has been traded for the non-zero lattice
  temperature $\tilde T=T/\alpha^2$ in the scalar sector.}

In reality, we cannot just discard the extra terms denoted by
ellipsis in Eq.~(\ref{new-temp-deriv}), but we certainly have
freedom to choose the value $\tilde T=T/\alpha^2$ for the dual
temperature as long 
as the general duality condition~(\ref{duality-general}) is satisfied.
With this particular choice the exact dual gauge kernel becomes 
\begin{equation}
  \label{simp-dual-gauge-kernel}
  \tilde\K=-{\I\alpha P\over 2\pi} 
  {1\over\displaystyle \I x+{(1+\I x)}^{-1}}, 
\end{equation}
where we denoted $x= TP/2\pi\alpha$.  The exact kernel $\tilde\K$
differs from the truncated dual kernel 
$$ %
-{\alpha\over2\pi}\e P_{\rho}\equiv -{\alpha \over 2\pi}[\I]_{\mu\nu}P
$$ %
only by corrections of the higher order $\tilde{A}{\cal
  O}(P^3)\tilde{A}$, so that in the na\"{\i}ve continuum limit the
dual gauge field $\tilde{A}$ does not have any $F^2$ term or
associated dynamics. 

It is tempting to propose that, as long as these corrections are
small, as terms of higher order in momenta they should be irrelevant
and the scaling properties of both the truncated and the exact forms
of dual theory should be equivalent with the corresponding critical
indices equal.  Then, using the previously established mapping between
theories that are dual as specified by Eq.~(\ref{duality-general}),
one could conclude that the two models with the gauge kernels of the
form~(\ref{non-local-cs}) and the corresponding parameters
$(T,\alpha)$ and $(T/\alpha^2, -1/\alpha)$ should be also equivalent
at large scales.  The first statement is, however, not so obvious
because the Chern-Simons theories are intrinsically non-local and the
critical dimensions of different operators may substantially differ
from their classical values.  We shall prove the legitimacy of this
procedure in the Section~\ref{sec:hierarchy}.

Along with the dimensionless parameter $\alpha$ we carried out the
transformation of the classically irrelevant dimensionful lattice
temperature $T$.  Clearly, since the values of the lattice momentum
$P$ are limited from the above, the small values of the ratio 
$T/|\alpha|=\tilde T/|\tilde \alpha|$ ensure the smallness of $x$, so
that the effect of higher order irrelevant terms is negligible and the 
bare value of the lattice temperature $T$ should be meaningful.
Remarkably, in the presence of the Chern-Simons term the duality
transformation does not lead to
reversion\cite{Kiometzis-93,Kiometzis-94} of the temperature axis
characteristic of the duality between the superconductor and the
superfluid.  In our case only the 
Chern-Simons coupling constant is inverted while the scalar sector
may remain in the strong coupling regime. 

The physical properties of excitations in the generalized x-y model
are revealed by their coupling with external fields in 2+1 dimensions.
Luckily, the introduction of the additional external gauge field $eA_0$
does not influence the exact procedure of the duality transformation;
now it results in the partition function~(\ref{villain-xy-dual}) up to the
substitution $A_{n\mu}\rightarrow{A}_{n\mu}+e{A_0}_{n\mu}$.
Integration over the original gauge field $A$ and proper shift of the
dual gauge field $\tilde{A}$ results in the frozen
model~(\ref{dual-zero-temp}) with additional minimal coupling to the
external field 
$$ %
-{e\over\alpha} {1\over 1+\I x}A_0,\quad 
x={TP\over2\pi\alpha}
$$ %
and the constant term 
\begin{equation}
  \label{external-coupling}
  {e^2\over4\pi\alpha} A_0(-k){\I P\over1+{\I x}}A_0(k)
\end{equation}
in the exponent.  These two terms remain intact in reparametrized
models with non-zero lattice temperatures.

It is clear that the model describes particles with the charge
$\tilde e=-e/\alpha$, while the momentum-dependent term $1+{\I x}$
in the denominator of the coupling can be regarded as a formfactor 
indicating the finite size of excitations and the presence of the
Magnus force.  The term~(\ref{external-coupling}) dependent only on
the external gauge field shows that the duality transformation
separated the condensate 
with the total charge density $-e^2\bd\times
A_0/2\pi\alpha=-e^2B_0/2\pi\alpha$ 
determined by the external magnetic field $B_0$.  More specifically,
if the original theory had the average charge density $\rho$, 
the total charge density of excitations in the dual representation is
given by 
\begin{equation}\label{rho-changed}
\tilde\rho=\rho-{e^2 B_0\over2\pi\alpha}.
\end{equation}
One flux quantum of
the magnetic field binds the charge $e/\alpha$, or exactly one
quasiparticle in the dual representation.  Since the duality
translates the original quasiparticles with charge $e$ into vortices
with unit vorticity, this condensate can be also
interpreted as the average vorticity $1/\alpha$ per quasiparticle in 
the dual model. 

Certainly, the external magnetic field does not at all simplify the
lattice model: even the one-particle Hofstadter spectrum in the
magnetic field is very complicated.  However, if both the magnetic
field and the particle density are small at the scale of the lattice,
only the filling factors of electrons $\nu=2\pi\rho/e^2B$ and
quasiparticles $\tilde\nu=2\pi\tilde\rho/\tilde e^2 B$ are important.
In this case the equation~(\ref{rho-changed}) becomes
\begin{equation}
  \label{nu-changed}
  {\tilde\nu}=\alpha(\alpha\nu-1). 
\end{equation}

\section{Periodicity in the Chern-Simons coupling}
\label{sec:periodicity}
Non-relativistic continuum models display the periodicity in the
Chern-Simons coupling because geometrically this coupling is just the
linking number between the trajectories of quasiparticles.  As long as
this linking number remains integer, equal increments in the
Chern-Simons  coefficient result in the total shift of the
phase of the partition function by integer multiples of $2\pi$.  This
is always so when quasiparticles avoid each other---for
example, if they are fermions or if they have strong hard core repulsion.
Although the x-y model {\em is\/} already in the limit
of point-like strong repulsion as viewed from the
corresponding continuum model, the current lines can intersect or even
join each other and special effort is needed to define the
Chern-Simons action periodic in its coupling.  
In this section we find such definition and
use it to derive the exact transformation analogous to the flux
attachment transformation in the continuum, valid for an arbitrary
form of the lattice Chern-Simons action.

The current of quasiparticles in the x-y model is defined as the field
canonically conjugated with the gradients of phases
$\fd_{\mu}\theta_n$.  Since the partition function is periodic in
these phases, they are cyclic variables and the current is
integer-valued.  As one would expect, the current operator is not
diagonal in terms of the phases~$\theta_n$; it is more convenient to
deal with currents in the special current representation.  In the
Villain approximation the explicit form of this representation can be
obtained with the inverse of the Poisson summation
formula~(\ref{summation-formula}).  The integration over phases
$\theta_n$ ensures the conservation of the current $M_{n\mu}$, and the
 partition function (\ref{villain-xy},\ref{gauge-part}) rewritten in
 terms of this current becomes 
\widetext
\begin{equation}
  \label{real-current-representation}
  Z=\sum_M\delta_{0,\bd_{\mu} M_{n\mu}}
  \int_{dA}\!\!\! e^{\textstyle-TM_{n\mu}^2/2 -iM_{n\mu}A_{n\mu}
    -{A_{n\mu}\K_{nn'}^{\mu\nu} A_{n'\nu}/2}  }.
\end{equation}
\narrowtext
It is important to emphasize that the transformation to the integer
current representation~(\ref{real-current-representation}) did
not change the gauge field $A$ in any way, so that the gauge kernel
$\K_{nn'}^{\mu\nu}$ in equations~(\ref{gauge-part}) and
(\ref{real-current-representation}) is exactly the same. 

The linking number between lattice current flow lines does not have
a natural geometrical meaning at intersection points and therefore
cannot be uniquely defined to fit all intuitive requirements.  There
is, however, no difficulty in defining the linking number for any two
conserved fields $M_{n\mu}$ and $N_{n\mu}$ determined on mutually dual
lattices since they never intersect.  Formally, this
coupling can be introduced via the solution $a^0$ of 
equations
\begin{equation}
  \label{aux-field-eqns}
  \e\fd_{\mu}{a^0}_{n\nu}= N_{n-\hat\rho\rho}, \quad
  \bd_{\nu}{a^0}_{n\nu}=0
\end{equation}
as ${\cal N}=\sum{M}_{n\mu}{a^0}_{n\mu}$.  Since the field $N_{n\mu}$
is defined on the links of the dual lattice (or at the plaquettes of
the primary lattice,) the auxiliary field ${a^0}_{n\nu}$ is defined on
the links of the primary lattice, and the summation is well defined.
One can prove that this definition is indeed the integer-valued linking
number by rewriting the conserved current $M_{n\mu}$ as a
superposition of some number $n_{\ell}$ of directed loops $L^{i}$,
$i=1,\ldots,n_{\ell}$ carrying unit current each.  This splits the
expression for ${\cal N}$ into a number of sums over independent
closed loops and by the Stokes theorem each of them is exactly equal
to the (integer) flux of current $N$ through the surface delimited by
the corresponding loop; obviously the total $\cal N$ is the
integer-valued linking number.  

Now let us define the regularized integer {\em self\/}-linking number
of the integer-valued current $M$ in exactly the same way but with
additional identification $M_{n\mu}\equiv N_{n\mu}$.  This identifies 
neighboring parallel links of the original and dual lattices mutually
displaced in the direction (1,1,1) by half a period.  This
displacement, uniquely determining the integer-valued regularized
linking number, is revealed in Fourier representation, where the
auxiliary equations~(\ref{aux-field-eqns}) take the form
\begin{eqnarray*}
  \e \left(e^{ik_{\mu}}-1\right) e^{ik_{\nu}/2} {a^0}_{\nu}({\bf k})
  &=&e^{-ik_{\rho}/2}M_{\rho}({\bf k}), \\
  \left(1-e^{-ik_{\nu}}\right){a^0}_{\nu}({\bf k})&=&0, 
\end{eqnarray*}
(no summation in $\rho$!) or, somewhat more symmetrically,
\begin{equation}
  \label{aux-eqns-fourrier}
  \e P_{\mu} {a^0}_{\nu} =i e^{-i(k_{x}+k_{y}+k_{z})/2}M_{\rho}, \quad
  P_{\nu}{a^0}_{\nu}({\bf k})=0.
\end{equation}
Solving these equations with respect to auxiliary field ${a^0}$, we
obtain the regularized self-linking number in Fourier representation
$$ %
{\cal N}={\rm Tr}\,{a^0}_{\mu}(-{\bf k}) M_{\mu}({\bf k}) %
=i{\rm Tr}\,\e M_{\mu}(-{\bf k}){Q_0 P_{\nu}\over P^2} %
M_{\rho}({\bf k}),
$$ %
where the original exponent $\exp{i/2\sum_{\mu}k_{\mu}}$
in~(\ref{aux-eqns-fourrier}) was replaced with $Q_0\equiv Q_0({\bf 
  k})=\cos{\sum_{\mu}k_{\mu}/2}$ allowing for the symmetry of the 
total sum.  The exponent of the linking number $\cal N$ can be 
further transformed to form a gauge coupling
\begin{eqnarray}
  \label{add-on-action}
  \lefteqn{ \exp{\textstyle -i\pi\alpha{\cal N}}\longrightarrow}
  &\relax& \\  \nonumber 
  &\relax& \qquad\int_{da}{ \exp{\sum_k
      -iM_{\mu}(-k){a}_{\mu}(k) + a(-k) {\I P\over4\pi\alpha Q_0}a(k)
      }},
\end{eqnarray}
with new transverse fluctuating field $a$ introduced as the
Hubbard-Stratonovich field.  Clearly, the gauge coupling has exactly
the form of that in~(\ref{real-current-representation}) with the gauge
kernel 
\begin{equation}
  \label{cs-kernel-local}
  \K_{\mu\nu}({\bf k})=\e{P_{\rho}\over2\pi\alpha Q_0}. 
\end{equation}
The momentum expansion of this kernel starts with linear antisymmetric
term and has no quadratic in momenta part; therefore, it corresponds
to the Chern-Simons term in the long-distance limit.  The apparent
singularity of this gauge kernel at the plane
$k_x+k_y+k_z=\pi$ does not lead to any divergences but merely
suppresses fluctuations of the field $a$ in the vicinity of this
plane.  One can totally avoid introducing somewhat unpleasant division
by zero by 
considering periodic lattices with an odd number of lattice nodes in
each direction, so that at any finite system size the singularity of the
kernel is never reached.  This also eliminates the problem of zero
functional denominator arising from integration over the transverse
part of the new gauge field $a$ in~(\ref{add-on-action}).

Consider the model~(\ref{real-current-representation}) with the
specific form of the gauge kernel~(\ref{cs-kernel-local}).  This
partition function is convergent at all positive lattice temperatures
$T$ even though the second term can be zero for certain
configurations.  One can use the Poisson summation formula to rewrite
the model in the form~(\ref{villain-xy},\ref{gauge-part}) depending only on
the phases~$\theta$.  By construction, this model is a periodic
function of $\alpha$ with the period $\Delta\alpha=2$, so that any two
models with the couplings related by
\begin{equation}\label{periodicity}
  \alpha\rightarrow\alpha+2m,\quad T\rightarrow T
\end{equation}
are absolutely equivalent to each other.  Definitely, this periodicity is
precisely the flux attachment symmetry as it was formulated for
non-relativistic systems.  It is important to mention that none of the
previously considered
definitions\cite{Frohlich-88,Luscher-89,Muller-90,Fradkin-89A,Fradkin-90}
of the lattice Chern-Simons term provide for this property; this is why
we needed to construct yet another form of the Chern-Simons coupling. 
The flux attachment transformation changes the properties of the
gauge field, but 
our main interest is to understand the effect of the fractional
statistics on the dynamics of the scalar field; we use the {\em
  auxiliary\/} 
Chern-Simons field only to get rid of the non-local interaction.

Even though one can construct other forms of the gauge coupling leading
to the same symmetry of the partition function, these forms are very
special; generic gauge couplings do not reveal the exact flux
attachment {\em symmetry}.
Moreover, it is easy to check that this property is not even preserved
by the duality transformation~(\ref{duality-general}).  Therefore, we
need to define the flux attachment {\em transformation\/} to 
display the same periodicity~(\ref{periodicity}) at least in the
continuum limit for some broad class of possible lattice
Chern-Simons terms.

In order to do this, let us introduce the trivial phase $2\pi m {\cal
  N}$ proportional to the integer-valued
self-linking number $\cal N$ {\em in addition\/} to the already
present in Eq.~(\ref{real-current-representation}) 
gauge coupling with the  
gauge kernel $\K$.  Clearly, this trivial phase does not change the
partition function 
and yet, after rewriting the introduced term as the additional gauge
coupling~(\ref{add-on-action}) and integrating away one of the gauge
fields, we obtain the same model but with the gauge kernel
\begin{equation}
  \label{flux-attached-kernel}
  \K'=\left(\displaystyle\K^{-1}-{4\pi m \I \,Q_0\over P}\right)^{-1}.
\end{equation}
The resulting model is rigorously equivalent to the original model
with the gauge kernel $\K$ and it is convergent at any 
non-zero lattice temperature $T$.

The long-range properties of the flux-attached gauge
kernel~(\ref{flux-attached-kernel}) depend on the properties of the
original kernel $\K$.  If $\K$ has the Chern-Simons form, {\em i.e.}\
its momentum expansion starts with the linear in momenta term 
$\K_{\mu\nu}\sim \e k_{\rho}/2\pi\alpha$, the expansion of the new
gauge kernel $\K'$ has the same form with the coefficient
$\alpha'=\alpha+2m$.  In the particularly simple case when the matrix
structure of the gauge kernel $\K_{\mu\nu}$ is the combination of
$\delta^{\rm t}_{\mu\nu}$ and $[\I]_{\mu\nu}$, we may further simplify
the flux attachment transformation by defining the formfactor $Q$
through $\K=\I P Q^{-1}/2\pi\alpha$.  Then the appropriately defined
formfactor $Q'$ for the new kernel $\K'$ is determined simply as the
linear combination
\begin{equation}
  \label{flux-attached-formfactor0}
  Q'={\alpha Q+2 m Q_0 \over \alpha+2m}.
\end{equation}
Earlier we adjusted the lattice temperature to obtain the purely
statistical Chern-Simons field without any 
dynamics in the na\"{\i}ve continuum limit, the appropriate lattice
gauge 
kernel having no quadratic terms in small momentum expansion.  This 
is equivalent to the expansion of the formfactor $Q(k)=1+{\cal
  O}(k^2)$, $k\rightarrow0$; obviously this property is preserved by
the equation~(\ref{flux-attached-formfactor0}). 

Before discussing the implications of this flux attachment
transformation on the universality of the phase transitions in
Chern-Simons models, let us reflect on the fact that the derived
expression~(\ref{flux-attached-kernel}) is an exact symmetry of any
lattice model with the integer-valued current, moreover, it is valid
not only for Villain models, but also for usual x-y models minimally
coupled with with some fluctuating gauge field.  If
the original model has no long-range 
interaction between the quasiparticles, it is in the universality
class of the x-y model, while the transformed model has a Chern-Simons
term with the gauge kernel~(\ref{cs-kernel-local}) and the coupling
$\alpha=2m$.  The duality transformation results in the model of the
same form with the lattice temperature $\tilde{T}=T/4m^2$ and
Chern-Simons coupling $\tilde{\alpha}=-1/2m$, the small momentum
expansion being valid as long as $TP\ll 4\pi m$.  The continuum limit
of this model is that of scalar particles coupled with the
even-denominator Chern-Simons field; yet this theory is rigorously 
equivalent to the original x-y model!

Completely different situation arises if we try to add such additional
Chern-Simons interaction to lattice superconductor described by the 
same model~(\ref{villain-xy},\ref{gauge-part}) but with {\em
  quadratic\/} in momenta gauge kernel $\K$.  Qualitatively, the
interaction introduced with the 
CS coupling is working independently of the original one.  In 
coordinate representation this additional interaction between 
current lines decays as $1/r$ at large distances; in order to have any
effect it should not be screened by the original interaction.
However, the current lines in normal state already have the magnetic 
interaction {\em logarithmic\/} at large distances, and the
additional coupling cannot have any effect on the long-distance
physics: it will not be visible at all at some finite scale. 

Although this statement appears to be rather evident, in some
cases the form of interaction at large distances may not be obvious.
For example, Schultka and Manousakis\cite{Schultka-94} considered the
3D x-y model with the Chern-Simons coupling introduced for {\em
  vortices\/} instead of the real charges.  Numerically, the phase
transition in this model is in the x-y universality class for all
values of the Chern-Simons coupling, and this was interpreted as the 
explicit proof of universality of phase transitions in the
Chern-Simons theories.  We do not agree with this interpretation:
clearly, the model dual to the x-y model is the frozen lattice
superconductor and the CS coupling between the vortices of the
original model is just the extra coupling between the currents of the
dual theory.  But the currents in the normal state of superconductor
already have the long-range 
magnetic interaction, the additional coupling is screened, and effectively 
the large distance behavior is merely that of the original x-y model. 
This statement can be obtained more rigorously from the small-momentum
expansion of the combined gauge kernel derived in the
Appendix~\ref{sect:manousakis}.  


\section{The Law of Corresponding States}
\label{sec:hierarchy}
For the Villain x-y model~(\ref{villain-xy},\ref{gauge-part}) with
fractional statistics $\alpha$ introduced by coupling to an auxiliary
gauge field with the kernel obeying
$\K^{\mu\nu}=\e k_{\rho}/2\pi\alpha +{\cal O}(k^3)$ at small momenta,
we found that the exact duality transformation~(\ref{duality-general},
$\tilde T=T/\alpha^2$) and flux attachment
transformation~(\ref{flux-attached-kernel}, $T'=T$) have very natural
and simple form
\begin{eqnarray*}
  &\tilde\alpha=-1/\alpha,&\quad \tilde T= T/\alpha^2\\ 
  &\alpha'=\alpha+2m, &\quad T'= T. 
\end{eqnarray*}
in terms of the Chern-Simons coefficient $\alpha$.
Although $\alpha$ fully describes the gauge
coupling only in the limit of small momenta, we would like to check
whether these relationships are meaningful by themselves at least for
some values of the lattice temperature $T$ and the Chern-Simons
coupling $\alpha$. 

In the continuum, the duality and flux attachment transformations can
be used to ``unwind'' the Chern-Simons 
coupling with the fractional parameter $\alpha_0=p_0/q_0$.  Shifting
$\alpha_0$ by an appropriate even integer number, we can always satisfy
inequalities 
$-1<\alpha_0'=\alpha_0+2m_0\le1$.  If $\alpha_0'=1$, or $\alpha_0'=0$,
we cannot proceed any further with our transformations and should
stop.  Otherwise the duality transformation yields
$\alpha_1=-1/\alpha_0'=p_1/q_1$ with new integer denominator
$0<q_1<q_0$.  Repeating such transformations, in a finite number of
steps $n<q_0$ we arrive at either $\alpha_n'=1$ or $\alpha_n'=0$ with
the final lattice temperature $T_n=q_0^2 T_0$ given simply by the
denominator $q_0$ of the original Chern-Simons coupling.  Since both
the shift and the duality transformation preserve the parity of the
sum of numerator and denominator of $\alpha$, it is clear that any
fraction with this sum odd ({\em i.e.}\ either numerator or
denominator is an even number), will lead to $\alpha_n=0$, and the
original fraction with both $p$ and $q$ odd will result in $\alpha_n=1$.

\begin{figure}[htbp]
    \leavevmode\centering
    \def\ifundefineddddd#1{\expandafter\ifx\csname#1\endcsname\relax}
    \ifundefineddddd{epsfbox}\relax\else
    \epsfxsize=0.6\textwidth
    \epsfbox[92 47 400 302]{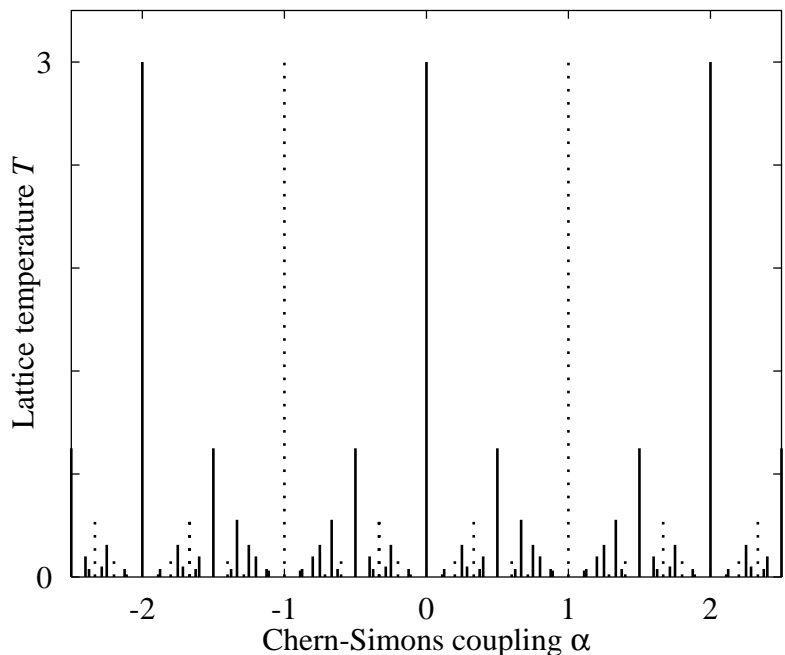}\fi
    \caption{The phase diagram of the Chern-Simons Villain x-y model.   
      Vertical lines represent ordered phases of different symmetry,
      corresponding to different levels of hierarchy.  The phases with
      the sum of numerator and denominator {\em odd\/} can be mapped
      to usual Villain x-y model, they are shown with solid lines.
      The phases with ``fermionic'' Chern-Simons coupling are only
      schematically shown with dotted lines since we know only the
      ratios of corresponding transition temperatures. 
      Unlike the case of usual superconductor,
      duality transformation does not invert the direction of the
      lattice temperature 
      axis, and the high-temperature phases are always disordered.}
    \label{fig:one}
\end{figure}

In the former case the final theory is just the pure x-y model
without any additional coupling.  Up to finite renormalization
correction, the phase transition temperature in this model is just
that in the usual Villain x-y model\cite{Kleinert-89}, $T_n\sim
T_{\rm  VXY}=3.03$.   We immediately obtain values 
\begin{equation}
  \label{high-level-temp}
  T_0\left(\alpha={p_0\over q_0}\right)
  \sim {T_{\rm VXY}\over q_0^2}
\end{equation}
of the phase transition temperatures in the Villain Chern-Simons
models with either numerator or denominator given by even numbers, as
plotted in Fig.~\ref{fig:one} with solid lines.   Similarly, once the phase
transition temperature in a single x-y model with ``fermionic''
Chern-Simons coupling $\alpha=1$ is known, one can use the same
relationship to obtain the approximate values of the phase transition
temperatures for all ``fermionic'' fractions with both numerator and
denominator odd.  

Let us rectify the outlined procedure by using the exact
duality~(\ref{duality-general}) and the flux
attachment~(\ref{flux-attached-kernel}) transformations at every step,
starting with the local theory with the gauge
kernel~(\ref{cs-kernel-local}) and the initial parameters $\alpha_0$
and $T_0$.  The partition function with this gauge kernel is periodic
with respect to the coupling $\alpha_0$, and we can safely assume that
$0<|\alpha_0|<1$.  The dual model has the lattice temperature
$T_1=T_0/\alpha_0^2>T_0$ and the gauge kernel
\begin{equation}
  \label{dual-local-gauge}
   \K_1\equiv\tilde\K_0={\I P\over2\pi\alpha_1}{1\over 
    \displaystyle {\I x_0}+\left(Q_0+\I x_0\right)^{-1}}
\end{equation}
where $x_0={PT_0/2\pi\alpha_0}$, $|x_0|\le T_0/\pi|\alpha_0|$ and the
the new 
coupling $\alpha_1=-1/\alpha_0$.  It is convenient to rewrite 
the new gauge kernel as
\begin{equation}
  \label{splitted-gauge-kernel}
  \K_1={\I P\over 2\pi\alpha_1 Q_1},
\end{equation}
by introducing the next level formfactor 
\begin{equation}
  \label{dual-formfactor}
  Q_1= \I x_0+{1\over Q_0+\I x_0}. 
\end{equation}
Similarly introduced formfactor of the CS coupling resulting from the
flux attachment transformation~(\ref{flux-attached-kernel}) can be
expressed merely as 
\begin{equation}
  \label{flux-attached-formfactor}
  Q_1'={\alpha_1 Q_1+2 m_1 Q_0 \over \alpha_1+2m_1}, 
\end{equation}
where $\alpha_1+2m_1=\alpha_1'$ is the new effective gauge coupling;
obviously, both transformations preserve the small momentum expansion
of the formfactor $Q(k)=1+{\cal O}(k^2)$.  
At the $n$-th step $\alpha_n$ becomes an integer and the universality
class of the resulting theory is determined by the parity of
$\alpha_n$.  

If this number is an even integer, the final
shift similar to~(\ref{flux-attached-formfactor}) results in the
gauge kernel {\em singular\/} at small momenta, and the long-distance
gauge interaction is suppressed.  The remaining short-range
interaction between the currents  perturbs the x-y critical point, and
one can prove that it 
is irrelevant in this point by simple power counting.  
As formally irrelevant, this additional interaction may result either
in some finite correction to the transition temperature, or it has to 
change the symmetry of the phase transition completely. 
In the former case, since the
models at every level of hierarchy are exactly equivalent to the
original Chern-Simons model with the fractional coupling $\alpha_0$,
we conclude that the phase transition in the original model is in the
same x-y universality class.

Similar arguments apply when the final $\alpha_n$ is an odd number
corresponding to the Fermi statistics of the quasiparticles.
Although we do not know the universality class of the phase transition
for Chern-Simons bosons with odd-integer-valued coupling, we can
claim that it should be the same for all original models in this class
unless the irrelevant terms drive it away from the critical point. 

So far we concentrated our attention on more pleasant possibility
that the irrelevant terms are too weak to change the symmetry of the
critical point and the phase transition does not change its
universality class or become the first order.  Let us try to analyze
the irrelevant terms more carefully to see what fractions are likely
to follow the universality scenario.  

We saw that the bare lattice temperatures of the transitions decrease
rapidly with the hierarchy level, therefore the temperature-dependent
convergence condition $x_k=PT_k/2\pi\alpha_k\ll1$ is not very
limiting.
There is, however, the momentum contribution from the formfactor
$Q_0=\cos\sum_{\mu}k_{\mu}/2$ of the local Chern-Simons
kernel~(\ref{cs-kernel-local}).  To 
investigate the extreme possible effect of this contribution, let us
write the sequence of hierarchical formfactors at zero lattice
temperature for the fraction 
\begin{equation}
  \alpha_0 %
  ={\displaystyle 1\over 2m_1 
    +\displaystyle { 1\over 2m_2+\ldots}},
\end{equation}
starting with the formfactor $Q_0$.  The first duality transformation
in the unwinding procedure results in $Q_1=1/Q_0$,
$\alpha_1=-1/\alpha_0$, and the flux attachment
transformation~(\ref{flux-attached-formfactor}) leads to the formfactor
\begin{equation}
  Q_1'={1\over Q_0}
  \left(1-2m_1\left(2m_2+{1\over2m_3+\ldots}\right){\cal P}^2\right),
\end{equation}
where 
\begin{equation}
  {\cal P}^2 =1-Q_0^2 =\sin^2\sum_{\mu} k_{\mu}/2
  \approx (\sum_{\mu} k_{\mu})^2/4
\end{equation}
vanishes at the origin.  Clearly, the coefficient in front of the
quadratic in momenta part increases with the increased levels of
hierarchy, and eventually it may become the main driving term of the
phase transition in the system. 

Making yet another duality transformation we obtain $Q_2=1/Q_1'$; the
new zero-temperature kernel has zero surface at finite momenta
determined by the equation %
${\cal P}^2=1/2m_1(2m_2+({1/2m_3+\ldots}))<1/2m_1$. %
Although the divergence does not occur at any finite lattice
temperature, the system develops a soft mode much closer to the origin
then the location of the original soft plane in the middle of the
Brilluen zone ${\cal P}^2=1$.  The exact location and the orientation
of this soft surface depends on the details of the selected regularization
procedure (in our case it was determined by the chosen form of
the local kernel~(\ref{cs-kernel-local})), 
but its existence is probably unavoidable as long as we
want to define the  integer-valued linking numbers.  The fluctuations
in the vicinity of this mode will grow with the hierarchy level and,
again, are capable of destroying the second order phase transition.

We see that the ``unwinding'' procedure creates additional
instabilities of the strength increasing with the level of hierarchy
and, generally, with the denominator of the statistical coupling.
Therefore, the
phase transitions in the system of particles with fractional
statistics $\alpha=p/q$ are expected to be universal only for small
enough denominators $q$. 

One could argue that instead of starting with the Chern-Simons bosons
with fractional statistics, we could have followed the usual direction
of the hierarchy sequence and started with the x-y model or Chern-Simons
bosons with odd ``fermionic'' statistical coupling.  In this case the
sequence of exact transformations also results in non-local
Chern-Simons model with fractional coupling, and the truncated local
model with the 
same Chern-Simons coefficient differs from the non-local one by
classically irrelevant terms.  Again, the two models should be in the
same universality class as long as the phase transitions remain of the
second order.  This procedure has the advantage 
that one can construct the sequence of lattice models
without any apparent divergences.  Indeed, starting with the
$\alpha_0=2m_0$, the duality~(\ref{dual-formfactor}) and the flux
attachment~(\ref{flux-attached-formfactor}) transformations result in
the formfactor
\begin{equation}
  \label{forward-hierarchy}
  Q_1'={1\over Q_0}\left(1-{2m_1 {\cal P}^2\over \displaystyle
      2m_1+{1\over 2m_0}}\right), 
\end{equation}
the coefficient in front of ${\cal P}^2$ here is less then one and the
original convergence condition $|k|\ll 1$ is preserved.
Unfortunately, this
argument also fails to prove the universality at high hierarchy
levels: the perturbation is not small compared to the 
relevant scale given by the vanishing lattice
temperature~(\ref{high-level-temp}).  
\section*{Conclusions}
\label{sec:conclusions}
In this paper we studied a relativistic version of the
Chern-Simons-Landau-Ginzburg theory of bosons in the limit of strong
coupling.  We used the lattice representation of this model in terms
of Villain x-y model minimally coupled with the Chern-Simons gauge
field to access the strong coupling limit without the perturbation
theory.  This model has duality and flux attachment symmetries in the
long wave length limit, {\it i.e.}\ these symmetries are accurate up
to irrelevant cubic and higher order derivative terms.  There is no
single lattice model that can obey both symmetries exactly, but we
construct algebraically exact non-local duality and flux attachment
{\em transformations} corresponding to these symmetries in the
continuum limit.  These non-local transformations were used to to show
that there are only two universality classes in this model, one
corresponding to the pure x-y transition, and another corresponding to
``fermionic'' x-y transition, or the transition in the x-y model with
the Chern-Simons coefficient $\alpha=1$.   
The value of such an investigation is to establish a theoretical model
in which the universality in a CSLG theory can be proven beyond the
bounds of perturbation theory.

Although it may be possible to construct an experimental system that
would be described by such relativistically invariant model, this
model is not in the same universality class as the experimental
quantum Hall systems, where the relativistic symmetry is broken by
non-zero charge density, external magnetic field and the disorder
potential.  Theoretically, these effects can be incorporated in the
lattice model as additional external gauge fields without breaking the
exactness of the performed hierarchy transformations, but the extra
fields lower the symmetry of the problem.  Now every phase is
characterized by the non-trivially transforming
filling factor $\nu$ in addition to the original Chern-Simons
coefficient $\alpha$, and the exact mapping between different quantum
Hall states is absent.

This is not surprising, since the mapping preserves the information
about the exact configuration of disorder.  We believe, that the
disorder averaging with proper identification of 
the relevant degrees of freedom will increase the symmetry and reveal
the universality of phase transitions in this model.  
Qualitatively, we saw that the Chern-Simons interaction has no effect
if it is screened by some other independent long-range force.
Two-dimensional time-independent random scalar potential may be
treated as the interaction of infinite range in time direction;
apparently this interaction is strong enough to suppress or modify the
effect of the statistical coupling on localization phase transitions.

Our model also sheds some light on the important question of
existence of high-order hierarchy states in the quantum Hall effect.
Since the duality and periodicity transformations can simultaneously
become symmetries of the theory only in the long-distance limit, the
accumulation of extra terms, irrelevant near the critical point,
will eventually render the phase transition of the first order or
change its symmetry.  

Further research on this class of models should be concentrated on
understanding the models with several species of the scalar fields.
Clear understanding of the model with finite number of components is
necessary to apply the replica trick, which is the only possible way
to perform disorder averaging in the presence of interaction. 

Another perspective direction is to understand the hierarchical
picture within the rigorous approach by Chen and
Itoi\cite{Chen-94,Chen-95}.  Starting from the representations of the
$2+1$ dimensional Lorenz group, they established the exact
relationship of the spin of particles with the coefficient of their
Chern-Simons coupling, namely that all theories of particles with spin
$s$ and Chern-Simons coupling $\alpha$ describe the same irreducible
representation of the Lorenz group with the fractional spin
$s+\alpha/2$.  Although nominally the action of the particles with the
spin $s$ has $2s+1$ components, in $2+1$ dimensions the equations of
motion for massive particles leave only one independent component, and
the long-range properties of all particles with the same $\alpha$ but
the spins $s$ differing by an integer should be similar.  This can be
considered as the rigorous definition of the flux attachment
transformation for hard-core relativistic particles with Chern-Simons
interaction.

While finishing this work we learned that a similar model is studied
by Fradkin and Kivelson\cite{Fradkin-95-unpublished}.  They also use
duality and periodicity transformations to argue that phase
transitions in three-dimensional lattice-regularized models are
universal, although they are less specific about the form of the
Chern-Simons coupling at short distances.  In addition to the behavior
considered in this paper, Fradkin and Kivelson consider a model with
additional dimensionless non-local interaction between the
integer-valued currents, corresponding to the dissipative conductivity
in the quantum Hall effect.  As in Refs.\cite{Kivelson-92,Lutken-93},
this leads to the phase diagram with multiple self-dual fixed points.
Perturbatively, such interaction of finite strength can be generated
near the phase transition point\cite{Pryadko-94}, and the accumulation
of higher-order irrelevant terms considered in the present work can in
principle imply the change of the symmetry of the second order phase
transition instead of the first order phase transition.  

\section*{Acknowledgments}
We would like to thank E.~Fradkin, S.~Kivelson, \hbox{D.-H.}~Lee and
C.~A.~L\"{u}tken for stimulating discussions.  L.~P.~is grateful to
the IBM graduate fellowship program for funding his research
assistantship at Stanford University.

\appendix

\section{The alternative derivation of the finite temperature dual
  model}
\label{duality-derivation}
Instead of using the previously established symmetry
property~(\ref{equivalence-general}), we could have arrived directly
at the dual action with non-zero lattice temperature by equivalent
transformations of the gauge field in the exponent.  Indeed,
averaging the equation~(\ref{vort-sum}) in the original gauge field
$A_{n\mu}$ produces the kinetic term~(\ref{zero-dual-kernel}) for the
dual gauge field $a$, and one can further rewrite the exponent in
terms of new auxiliary transverse field $\underline b$ as 
$$
i a M -{1\over2}a\tilde\K_0 a
\rightarrow i a M -ia \underline{b} -
{1\over2}\underline{b}\tilde\K_0^{-1} \underline{b}. %
$$
Now one can add and subtract the term $\tilde T\underline{b}^2/2$
with the arbitrary constant $\tilde T$, so that the introduction of
yet another transverse field $\underline{c}$ to dispatch off the part
$\underline{b}\left(\tilde\K_0^{-1}-\tilde{T}\right)\underline{b}/2$
of the obtained gauge kernel results in the exponent
$$ %
i a M -ia \underline{b}+i\underline{b}\underline{c}
-{\tilde{T}\over2}\underline{b}^2    
-{1\over2}\underline{c}
\left(\tilde\K_0^{-1}-\tilde{T}\right)^{-1}\underline{c}.
$$ %
This expression is linear in the field $a$ and an easy integration
yields the constraint $\underline{b}=M$; the subsequent integration in
$\underline{b}$ trivially yields
$$ %
i M \underline{c}- {\tilde{T}\over2}M^2
-{1\over2}\underline{c}\left(\tilde\K_0^{-1}-\tilde{T}\right)^{-1}
\underline{c}.
$$ %
With the transverse integer-valued current $M$ this is precisely the
exponent of expression~(\ref{zero-dual-currents}) with the gauge
kernel $\tilde\K=\left(\tilde\K_0^{-1}-\tilde{T}\right)^{-1}$ and the
new lattice temperature $\tilde{T}>0$ replacing the artificially
introduced infinitesimal variable $t$.  Now the summation
formula~(\ref{summation-formula}) and subsequent rescaling of the new
gauge field lead directly to the dual model of the
form~(\ref{villain-xy},\ref{gauge-part}) with new finite temperature
$\tilde T$ and the gauge kernel $\tilde\K$
satisfying~(\ref{duality-general}).

The derived mapping is a generalization of the well-known duality in
three dimensions\cite{Peskin-78}.  It can be understood as the duality
between vortices and monopoles in three spatial dimensions, or the
charge---vortex duality in $2+1$ dimensional systems.  Indeed, the
average in the presence of an arbitrary number of vortex-antivortex
pairs introduced by the integer vortex strength $L_n$
\begin{equation}
  \label{vortices}
  \left\langle \exp{i\sum_n L_n \theta_n }\right\rangle
  \equiv\left\langle \exp{i\int\dk L_{-k} \theta_k }\right\rangle 
\end{equation}
with zero total vorticity $\sum_n L_n=0$ results in the
constraint
\begin{equation}
  \int {d\theta_n\over2\pi} e^{\textstyle i L_n
    \theta_n-i\theta_n\bd_{\mu}b_{n\mu}}=
  \delta\left(\bd_{\mu}b_{n\mu}-L_n\right)
\end{equation}
instead of~(\ref{conservation-of-b}).  Evidently, the integers $L_n$
serve as the sources for the field $b$, or the monopoles with
appropriate quantized charges.  The values of these charges are not
affected by any single-valued transformation of the vector potential,
therefore the statement of equivalence generally holds for any two
models related by Eq.~(\ref{duality-general}).  Similarly, the dual
transformation of the original
model~(\ref{villain-xy},\ref{gauge-part}) in the presence of monopoles
introduced by appropriate multivalued external vector potential
results in vortices of the form~(\ref{vortices}) located exactly in
the original positions of the monopoles.

Unfortunately, any averages involving the gauge fields, particularly
the averages of gauge-invariant currents, do not have simple dual
representation since the fluctuating gauge fields change upon
reparametrization.  Therefore, only scalar sectors of arbitrary models
of the form 
(\ref{villain-xy},\ref{gauge-part}) related by the generalized duality
transformation~(\ref{duality-general}) are equivalent to each other.

\section{CS coupling between the vortices of x-y model: detailed
  analysis} 
\label{sect:manousakis}

Here we analyze the model numerically considered by Schultka and
Manousakis\cite{Schultka-94}: the pure x-y model with the additional
Chern-Simons interaction between the {\em vortices}, introduced to
study the universality of transitions in a system of particles with
fractional statistics.  The original model is defined simultaneously
in terms of original phases $\theta_n$ and the vorticity $M_{n\mu}$.
Working in the Villain approximation, let us perform the exact duality
transformation on the scalar sector and express the model entirely in
terms of dual variables.  The duality transformation results in the
model of the form~(\ref{zero-dual-currents}) with the additional
coupling of the vorticity $M$ to the second gauge field $A$ with the
Chern-Simons kinetic term.  Shifting the gauge field 
$$ %
 A_{n\mu}\rightarrow A_{n\mu}- {a}_{n\mu}
$$ %
and integrating away the field ${a}$ we obtain the gauge kernel of the
combined interaction in the form
\begin{eqnarray}
  \label{converted-pure-xy}
  \K&=&{\I P\over2\pi\tilde\alpha Q}\left( 1-{1\over1-\I T\tilde\alpha Q
      P/2\pi}
  \right) \\ %
  &=& {T P^2\over4\pi^2}+{\cal O}(P^3), \qquad\tilde\alpha T P Q\ll
  2\pi,\nonumber
\end{eqnarray}
instead of $TP^2/4\pi^2$  present without any gauge coupling; the
statistical coupling of the vortices does not change the long-range
properties of the model at all!

This is precisely the result of numerical study\cite{Schultka-94};
from our analysis it is clear that the additional Chern-Simons
interaction between the vortices gets completely screened by the
Coulomb interaction already present in the dual model.  Indeed, the
original vortices of the x-y model map to charges in superconductor,
which is the correct dual model.  The charges have the Coulomb
interaction decaying like $1/r$ at large distances, while the
additional CS interaction falls off like $1/r^2$; there always exists
some range above which the usual Coulomb force dominates.  It is
because of the existence of this finite range the asymptotic form is
correct only in the restricted range of momenta as specified
in~(\ref{converted-pure-xy}).  

This phenomenon can also be understood in terms of the original x-y
model: the vortices in superfluid are not really local objects because
of their long-range phase structure, therefore the additional
Chern-Simons interaction does not do anything at large enough
distances.

It is important to emphasize that the previous analysis, including the
equation~(\ref{converted-pure-xy}), holds for x-y model coupled to a
Chern-Simons gauge field via the dual current only.  If the long-range
structure around the vortices is already broken by some mechanism, the
addition of the Chern-Simons interaction between vortices does change
the parameters or even the qualitative behavior of the model.  In this
case the long-range phase structure of the vortices is destroyed (in
the dual representation---the charges are screened) and the additional
statistical coupling is the main interaction at large distances.
\widetext


\begin{thebibliography}{88}
\bibitem{Note-1}
In three dimensional Euclidean theories the lattice temperature plays the role
  of an inverse coupling constant, it has nothing to do with the real
  temperature which is equal to zero in formally equivalent $2+1$ dimensional
  models.
\end{thebibliography}
\end{document}